\documentclass{article}

\usepackage{PRIMEarxiv}

\usepackage[utf8]{inputenc} 
\usepackage[T1]{fontenc}    
\usepackage{hyperref}       
\usepackage{url}            
\usepackage{booktabs}       
\usepackage{amsfonts}       
\usepackage{nicefrac}       
\usepackage{microtype}      
\usepackage{lipsum}
\usepackage{fancyhdr}       
\usepackage{array}
\usepackage{natbib}
\usepackage{pdfpages}
\usepackage{amsmath}
\usepackage{graphicx}       
\graphicspath{{media/}}     

\title{Algorithmic Tradeoffs, Applied NLP, and the State-of-the-Art Fallacy}

\author{
  AJ Alvero \\
  Cornell University \\
  \texttt{ajalvero@cornell.edu} \\
   \And
  Ruohong Dong \\
  University of Arizona \\
  \texttt{ruohongdong@arizona.edu} \\
  \And
  Klint Kanopka \\
  New York University \\
  \texttt{klint.kanopka@nyu.edu} \\
  \And
  David Lang \\
  Stanford University \\
  \texttt{(dnlang86@stanford.edu}
}

\begin{document}
\maketitle

\begin{abstract}
Computational sociology is growing in popularity, yet the analytic tools employed differ widely in power, transparency, and interpretability. In computer science, methods gain popularity after surpassing benchmarks of predictive accuracy, becoming the “state of the art.” Computer scientists favor novelty and innovation for different reasons, but prioritizing technical prestige over methodological fit could unintentionally limit the scope of sociological inquiry. To illustrate, we focus on computational text analysis and revisit a prior study of college admissions essays, comparing analyses with both older and newer methods. These methods vary in flexibility and opacity, allowing us to compare performance across distinct methodological regimes. We find that newer techniques did not outperform prior results in meaningful ways. We also find that using the current state of the art, generative AI and large language models, could introduce bias and confounding that is difficult to extricate. We therefore argue that sociological inquiry benefits from methodological pluralism that aligns analytic choices with theoretical and empirical questions. While we frame this  sociologically, scholars in other disciplines may confront what we call the “state-of-the-art fallacy”, the belief that the tool computer scientists deem to be the best will work across topics, domains, and questions.
\end{abstract}


\section*{Introduction}

In computational research, whether sociologically aligned or otherwise, analysts inevitably encounter a critical question: which modeling approach should I/we take? Model selection is a multifaceted process that requires weighing a variety of tradeoffs. In the machine learning literature, the well-known bias-variance tradeoff encapsulates the tension between statistical bias and variance \cite{hastie2009elements,james2013introduction}. From a more theoretical perspective, scholars emphasize the importance of balancing flexibility, accuracy, and parsimony in model selection as a way to ensure strong results that are still inferable \cite{xie1998comment}. In computational text analysis, a topic gaining prominence in sociology \cite{evans2016machine}, these tradeoffs take on unique forms. Modeling decisions in this area often reflect not only the balance between interpretability and flexibility but also entirely distinct approaches to representing text and generating variables. For instance, topic modeling generates representations of documents as mixtures of word co-occurrence clusters referred to as ``topics.'' In contrast, word embeddings focus on the relationships of individual words within a corpus, eschewing document-level representations in favor of high-dimensional word spaces. Beyond these two popular sets of methods, there are entire conferences and sub-fields dedicated to developing tools primarily geared for text analysis, making it difficult if not impossible to choose the one truly ``right'' method.





Therefore, as our social worlds become increasingly text-saturated and tools to analyze such data improve and popularize, scholars face the challenge of determining which computational methods best align with their research questions. Sociologists who advocate machine learning frameworks have articulated compelling arguments for the adoption of computational approaches as a means to leverage our modern data deluge \cite{borch_oxford_nodate,rahal2024rise}. For example, Verhagen \cite{verhagen2024incorporating} highlights how machine learning methods can enhance hypothesis testing and refine applied models. Similarly, Lundberg et al. \cite{lundberg2022researcher} discuss the potential of computational techniques to improve prediction and analysis in large datasets, offering insight into social phenomena that have the potential to surpass traditional statistical methods. However, these contributions often focus on general applications of computational techniques rather than on their specific implications for text analysis. 

Although sociologists have contributed to the development of computational tools adapted for social science, many of the widely adopted methods in machine learning and natural language processing (NLP) are developed by computer scientists whose priorities differ from those of social scientists. These tools are typically optimized for achieving maximum predictive accuracy on benchmark datasets rather than addressing the nuanced demands of social inquiry \cite{luitse2024ai,koch2024protoscienceepistemicmonoculturebenchmarking}. As Koch \& Peterson note, not a bottom-up cultural development in computer science but rather a heavily top-down phenomenon influenced by major funding sources like DARPA over decades. A result of this arrangement is that distinctions between algorithms that model text as mixtures of topics versus those that represent words in high-dimensional spaces or any other approach may seem less relevant in computer science than in sociology. Modeling decisions for computer scientists tend to be based on ``whatever scored highest on the latest benchmark'' \cite{donoho201750,koch2021reducedreusedrecycledlife} rather than ``whatever makes the most sense for the particular research question.'' These dynamics can limit computational social scientific research as well by pushing analysts to adopt the state of the art\footnote{Typically, ``state of the art'' is a noun and ``state-of-the-art'' is an adjective, a convention we follow in this article.} method as determined by computer science in all situations or even punish authors who use older methodological approaches. These dynamics also point to a potential ``file drawer problem'' \cite{rosenthal1979file} where computational and/or social scientists might be limiting the breadth of their approach in favor of whatever has become or currently is popular. Identifying this trend in the literature would therefore be difficult, though in this paper we take more of a conceptual approach and compare analytical frameworks comparing older methods with newer ones under the assumption some might have that the state of the art will generally work the best.

Sociologists adapting text analysis methods obviously implies that sociologists are receiving some kind of training in these techniques, be it through formal coursework or through semi-formal means like workshops and focused trainings (e.g., ICPSR). The majority of the top sociology programs in the U.S. offer some kind of graduate focused course in computational methods and analysis (21 of the top 31; see Table \hyperref[tab:comp_soc_training]{A1} for a breakdown based on publicly available information on department websites). However, slightly less than one third of them (10/31) offer curriculum specifically about computational text analysis, suggesting that most of the formal training of sociologists in using NLP to analyze text is happening outside of sociology departments. Other social science departments also active in computational text analysis, such as psychology and political science, probably take on some of that burden by allowing students outside their respective departments enroll in their courses (one such example is noted in {A1}). But beyond these instances, it seems likely that the current cohorts of graduate students who are doing a lot of work energizing computational sociology are also forced to learn how to actually do the work from computer science and aligned programs, workshops, and/or internet based trainings outside of their home discipline. In principle, learning from computer scientists or experts from workshops is not a problem. However, machine learning sub-fields in computer science tend to create and then follow hype trends very closely, sometimes to their detriment \cite{lipton2019troubling}. And the experts (along with computer scientists offering formal training) putting out training materials online or leading workshops probably have the skills and knowledge to identify issues when a particular method is not working well, something that students might find difficult to emulate themselves. Even beyond the issues related to technical training, crossing into disciplinary terrains could influence theoretical considerations, interpretation of results, and the very nature of the work. These trends have cascading effects as well, such as peer review and norm setting for the field. 

We term the belief that whatever the computer scientists label the state of the art will work the best for social scientific inquiry involving text the ``state-of-the-art fallacy'' and argue that we should be cautious in their assumptions about which methods will perform ``best.'' The nature of computer science research and its respective (high) status help contribute to this phenomenon. To illustrate our point, we compare analyses on the same dataset using both recent, state-of-the-art methods and older approaches on a dataset of college admissions essays (n = 238,892) written before the release of ChatGPT. In prior analyses, it was found that relatively interpretable features generated from an older machine learning approach were strong predictors of applicant information through an also interpretable out-of-sample linear model. But these methods did not reflect the state-of-the-art methods at the time (mostly word embedding methods and more sophisticated predictive modeling). Even more recent developments in generative AI have supplanted these approaches to become the new state of the art, a trend we also explore in a second study in this paper. Unlike widely used methods in sociology and other social sciences like linear regression, NLP encompasses a vast array of approaches and tools. Without a clear theoretical framework for model selection, scholars risk falling into the state-of-the-art fallacy. Our paper also makes a strong though implicit argument about the need for methodological breadth and depth when testing NLP tools, a point we come back to in the discussion.


New developments in NLP can offer exciting analytical opportunities, but social scientists should remain critical, employing methods that best suit their specific research questions and theoretical frameworks. Our purpose is not to provide an exhaustive description of NLP techniques, as many excellent resources already exist (see for example \cite{stoltz2024mapping,grimmer2022text}). Nor is our goal to create a shroud of suspicion when social scientists adopt new tools into their repertoire; in fact, there are analogous versions of our argument but focused on traditional tools like linear regression \cite{spirling2025good,abbott1988transcending}. Instead, we aim to help social scientists develop theoretical justifications for model selection that resist reliance on ``the latest and greatest'' while remaining open to innovative developments. Computer scientists have grappled with similar issues as well, such as the assertion that interpretable models should be used instead of black box models \cite[though importantly they do not suggest to stop research on black box models]{rudin2019stop}, and here we seek to extend this thinking into sociological research. And in sociology we also must consider the ways that black box models might be contaminated with poor quality data in ways that we will never be able to know, not to mention the fact that sociologists working in this space regularly have access to text datasets large enough for them to train their own models on their own data. Striking a balance between adopting computational innovation while staying true to social science inquiry will be essential for all social scientists moving forward. Doing so will ensure that methodological choices are driven by scholarly inquiry rather than technological trends. 


\section*{Background}
The messiness of choosing an NLP method stands in stark contrast to the broad usage on linear models in most other social science research. Linear modeling decisions also require some analytical decision making that can generate variation in the signs of coefficients or steepness of slopes \cite{salganik2020measuring,breznau2022observing}, but analogous decisions in NLP are much more consequential, multiplex, and encompassing. For example, while there is little argument in traditional quantitative analysis about preprocessing data, the same cannot be said for computational text analysis. Researchers must decide not just how to clean the data but also consider how they might define what constitutes a word; how to handle variation in things like spelling and the most/least frequently occurring words; and polysemy (words with multiple meanings). There are many heuristics to follow and consider but ultimately there is no truly right or wrong way to make these decisions, just like there is no truly right or wrong method to use. However, it is closer to the end of the NLP analytical pipeline where differences in decision-making and rationale have the most impact. 



This does not mean that the sociological literature is bereft of work relevant to this topic, however. Recent work by Macanovic \cite{macanovic2022text} and van Loon \cite{van2025metatheoretical,van2022three} have proposed frameworks that help researchers navigate the vast landscape of computational techniques. Macanovic emphasizes the integration of traditional theories with modern text-mining methods, while Van Loon’s typology categorizes approaches into term frequency analysis, document structure analysis, and semantic similarity analysis as a means to more clearly connect research questions to methodological justifications. These frameworks aim to guide researchers in selecting methods that align with their theoretical goals rather than defaulting to state-of-the-art tools. Both also highlight an important point about these tools: they are built for prediction and have become popular due to accuracy in different tasks. Similarly, Nelson’s computational grounded theory highlights the potential for combining machine learning with qualitative methods to better analyze textual data in the social sciences \cite{nelson2020computational}. We expand upon this literature by demonstrating that deviating from these sociologically inclined frameworks toward selecting the most sophisticated methods (as labelled by computer scientists) can lead to extreme disparities in outcomes.



While recent frameworks for computational social science have made strides in integrating data and theory, underlying tensions remain that are analogous to past meta-sociological theory. For example, C. Wright Mills warned that sociology which overemphasizes methodological sophistication and quantification may produce rich insights but unclear meaning that is detached from substantive theory, an argument he termed ``abstract empiricism'' \cite{mills1959sociological}. With big data and now AI, there is not only a stronger impulse for measurement and datafication in daily life and research but also a sophisticated technological ecosystem which pushes people in this direction. The data of course is not neutral, and we argue that it does not represent some previously unknowable truth, but as Espeland and Stevens note the abundance of data and subsequent patterns of quantification transform social inquiry in ways that obscure the context from which the data was generated and the normative reflections held by adherents to a higher truth in data \cite{espeland2008sociology}. Mills might have seen these developments and noted the stark increase of available data and tools to analyze them and the potential tradeoff of limited theoretical understanding and development. As sociologists continue to adapt tools from computer science without fully understanding the breadth of tools available or consider that the state of the art might not be the best tool, we run the risk of weakening our theoretical ambitions in favor of novelty in data and method.

Scholars have attended to these issues as they manifest in big data and computational methodology. Savage and Burrows' argued that there is a ``coming crisis'' in empirical sociology as the data deluge forces the discipline to engage with the ``politics of method'' and the potential reorientation of sociology away from its critical core \cite{savage2007coming}. boyd and Crawford's seminal paper likewise argues that big data does not always necessarily mean ``big'' insights and can actually lead to shallow understanding of complex social phenomena \cite{boyd2012critical}. The focus of this paper is not uniquely about big data, but the literature surrounding it has generated sharp insights that we draw upon in our arguments. A reasonable conclusion to draw from these perspectives could be that sociology might therefore turn away from computational tools and big data, but such a reading is not our goal. Rather, here we are connecting a body of literature pointing to potential limiting factors on theory but also pointing to how they might implicate methodology as well. The consequence is the same, however, as assuming that trends in computer science will be able to broadly address social scientific inquiry could introduce new types of limitations on the questions we ask. 

These issues also point to fundamental differences in epistemic cultures between computer science and sociology. Like all disciplinary fields, each has developed its own set of systems, tools, practices, and assumptions about how to produce and organize knowledge and research \cite{cetina1999epistemic}, including those which each might consider to be the ``state-of-the-art''. For sociologists, this usually looks like the reflexive connections between theory (or theoretical frameworks), a substantive topic or problem, and clear method alignment; this point has been previously established with computational sociology (and computational text analysis) as well \cite{bonikowski2022ends,edelmann2020computational}. The product of sociological work is also not quite built for market like computer science, where a research project might be spun out into a start-up or have some obvious patent potential. The ``state-of-the-art'' in sociology therefore looks less like something requiring hype and marketing from the start and more like the next link on a chain of understanding built from past connections to theory, data, and methods (even though those connections could sometimes be sharper; \cite{lundberg2021your}). Computer science has a much greater emphasis on a model's performance on some standardized benchmark, therefore enabling them to add some backing to their claims about which tool or approach is the ``best''. The very structure of this also denotes the cultural values of competitiveness deeply embedded into computer science, the winners of which can become epistemic gatekeepers about what counts as novel, innovative, and worthy of further adoption. 

With calling a method called ``state of the art'', it does more than its technical performance: people might also consider it is a form of linguistic capital \cite{bourdieu1991language}. Methods do not travel alone but also with its evaluative logics, and the logics in computer science are distinctive compared to sociology. By prioritizing methods, modeling performance and benchmark success rather than theoretical fits and sociological problems, sociologists risk reorganizing the values and internal hierarchies by other subjects. Therefore, newly imported methods carry an epistemic culture with them, such as benchmarks, hype and reshaping the internal hierarchies with sociology itself. Scholars who adopt the latest and most advanced models might be rewarded symbolic capital in the appraisal of their work, but those who rely on older ones risk being devalued, which is not coherent with the substances of their results. In this way, the state-of-the-art fallacy is not only methodological but also sociological in the way it reflects symbolic capital shifts across fields and the procedures of evaluation within sociology. 


When methods and techniques trickle out of computer science and into mainstream science (inclusive of sociology), it is likely to be the case that outsiders gain exposure though hype surrounding methods which lead any given benchmark (i.e., whatever is the current state of the art). And, as Evans and Aceves note, this also means that the tools being developed were borne from or influenced by the fairly narrow problem space of popular benchmark datasets \cite{evans2016machine}. Regardless of whether or not this particular paradigm works in the way intended, adopting computational tools without attending to the particularities of computer science's epistemic culture could have sociologists making inaccurate assumptions based on computer science trends about the connection between \textit{computational} methods, theory, and data. 

Importantly, the ``trading zone'' \cite{mcfarland2016sociology} between computational and social science is not unilateral. Some inside computer science have developed well thought out critiques of the structure of computer science as a discipline and the ways it can entrench bias and exclusion not just with respect to the science of computer science but also in the social harms from its outputs. These arguments often explicitly draw on sociological literature about these topics or even have sociologists as co-authors (e.g., \cite{paullada2021data}). There are however also examples of computer science theories (non-technical explanations as to why they believe a particular dataset or method worked in their case) that can also distort sociological inquiry. For example, a common explanation in computer science about the efficacy of word embeddings as a method was that they represented some kind of ``true'', ``latent'' meaning akin to a Platonic ideal of the meaning of a given word; this argument is often justified by referencing a theory from the 1950s called the ``distributional hypothesis'', where the meaning of a word is thought to be understood by examining which words appear around it; \cite{firth1957synopsis}). There is some truth to the basic premise in this theory, but computer scientists have argued that this explains why word embedding models are ``true'' (in a purely positivist sense) representations of meaning, meaning that variations in language unaligned with the model are implicitly ``wrong''. 

When computer scientists develop these theories, they are not rigorously tested using a variety of methods (including and especially causal frameworks). Rather, these perspectives are more akin to ``vibe theorizing'', not unlike the ``vibe coding'' which has become popular with generative AI\footnote{Vibe coding, a term popularized by computer scientist Andrej Karpathy, is currently defined as ``the practice of writing code, making web pages, or creating apps, by just telling an AI program what you want, and letting it create the product for you.'' See \url{https://www.merriam-webster.com/slang/vibe-coding}.}. Some sociologists have cautioned against overstepping assumptions lest we start to confuse the usefulness of NLP models with atheoretical hype generated from excited computer scientists \cite{kozlowski2019geometry,boutyline2025meaning,arseniev2024theoretical}. If such theoretical overfitting were to happen in sociology, computer scientists in general (not just those with critical leanings) might take that as theoretical validation for some aspect of not just word embeddings but any number of computational methods. Clearly, the stakes are bigger than just inadvisable use of contemporarily popular NLP methods. 


In the following sections, we explain the tradeoffs between modeling decisions and why these decisions matter to social scientific inquiry. We then explain how the ways that social scientists tend to use NLP for highly specified analyses (rather than as one piece of a larger system like much of NLP research) makes these tradeoffs highly consequential. Finally, we use these insights to explain our methodological contribution of the state-of-the-art fallacy. These sections also help set up an empirical demonstration where we show the ways that researchers might generate much weaker results if they follow the state-of-the-art rather than cast a broader methodological net. 

\section*{Algorithmic Tradeoffs}
Sociologists grapple with tradeoffs constantly: due to the impossibility of collecting data from each and every member of a social group or society, we use surveys and sampling procedures as a tradeoff to reduce costs while optimizing for inference. In computer science, particularly machine learning and NLP, these types of concerns matter much less if at all. Aside from notable subfields like interpretable, causal, and transparent machine learning, the primary goal of developing new algorithms is to maximize predictive accuracy on some shared benchmark or dataset. Put differently, sociological research willingly accepts some inefficiencies to preserve interpretability and inference, whereas machine learning research makes the opposite trade in favor of higher rankings on a leaderboard. As one famous example, Netflix had previously been using linear models to suggest movies to users and held an open contest for computer scientists to submit a more accurate method; the winning approach combined many different algorithms to create predictors, inference be damned. As computational methods developed under these more focused pretenses become more widely incorporated into the sociological toolkit, it is imperative that sociologists keep these somewhat disparate goals in mind.

For computer scientists, the relative opacity of a model is less relevant than its maximum performance. In fact, some of the most heralded advances in machine learning have been the most opaque but also the best at predictive tasks. On the supervised machine learning side, deep learning (a neural network model made up of dozens of  non‑linear layers trained with back‑propagation) surpassed previous approaches (and therefore received commensurate adulation, including the 2018 ACM Turing Award, the ‘Nobel Prize of computing’) despite inference into how or why the model made these predictions is essentially inscrutable. In NLP, similar trends emerged with the development of word embeddings (sometimes called ``word vectors''). They (first the word2vec approach followed by GloVe; \cite{mikolov2013distributed,pennington2014glove}) far surpassed the performance of other algorithms on shared benchmarks, regularly eclipsing previous methods on analogy tests, semantic similarity benchmarks, and downstream classification tasks by double‑digit margins. This trend has continued in the era of generative AI, where models are trained on massive corpora scraped from the internet using a deep learning approach and then fine tuned with human intervention (reinforcement learning with human feedback, or RLHF). The largest models contain trillions of parameters, making traditional inference impossible. Consequently, the very advances that motivate AI researchers often leave sociologists in a difficult position methodologically, because so far the gains in raw performance arrive with ever more opaque black boxes.

In contrast, the linear models commonly used in the social sciences—such as ordinary least squares (OLS) and logistic regression—prioritize interpretability and theoretical inference. What they lack in terms of flexibility, they more than make up for in transparency. Their current and historical popularity also make them extremely amenable to comparisons of results over time and across entire bodies of literature. And simply, these linear models allow researchers to examine directionality, magnitude, and statistical significance in a transparent and often theory-driven way. While more complex models have emerged over time in the statistical and machine learning literature, including support vector machines (SVMs), deep learning, XGBoost, and regularized methods like LASSO, they do not have the same straightforward path to interpretation as linear models. This is not always obvious though since these and other methods were designed as supervised machine learning tools where the final reported outcome is usually some kind of accuracy metric (e.g., F1 score, raw accuracy, area under the curve, etc.) that itself is easily communicated: the accuracy for model X was higher than model Y. But the underlying mechanisms as to the reasons why some predictor increased or decreased this metric are not as clear nor emphasized in computer science research. Considering the tendency to work with datasets that can be large, wide, or both, spending less time dissecting statistical patterns in each individual variable is an understandable outcome. On the industry side, Burrell notes this opacity can stem from things like intentional secrecy surrounding proprietary algorithms, a lack of technical fluency among users, and the intrinsic complexity of the models themselves \cite{burrell2016machine}. Even when transparency is not deliberately withheld, deep learning and ensemble methods are difficult to interpret due to their layered, non-linear structure. In response, machine learning researchers have increasingly focused on explainability, developing post-hoc tools and designing models that are more interpretable by construction \cite{borch2022toward}.  Still, these efforts are rooted in a different epistemic tradition where the tradeoff between predictive accuracy and interpretability is considered to be a good deal.

Such an approach though can go against social scientific norms and assumptions about not just the relationships between variables but even those related to the structure of a model, causal interpretation, and the perceived quality of a model. Issues like multicollinearity or endogeneity are not primary concerns like they are in the social sciences, but because they hinder social scientific understanding of explanatory mechanisms and structure they are taken more seriously. This does not mean that these tools are entirely incompatible with social science, though. Recent studies have adopted supervised machine learning frameworks to help answer questions ranging from racial threat narratives in US newspapers \cite{duxbury2023threatening} and describing gender inequalities in education \cite{mittleman2022intersecting}. There are also many papers where some kind of machine learning approach was used to scale up qualitative coding in some way \cite{stuhler2025codebooks,do2024augmented} and other that scope out the potential of otherwise inscrutable computational methods to be used as support for more traditional social scientific question answering and explanation \cite{koch2024primer,broska2024mixed}. But as many of these papers will also point out, adopting these methods will introduce new costs and tradeoffs to consider, especially as it relates back to transparency and interpretability. In our empirical analyses, we consider these perspectives but also challenge the assumption that more flexible (and less interpretable) models will usually or almost always get you stronger performance, even by the simplified metrics favored by computer science.

\section*{Applied NLP}

When computer scientists lead NLP projects, much of it will be with entire computational systems in mind. This includes not just the people designing the classifiers and algorithms which transform text data into numerical representations, but also computer scientists building entire platforms and databases, leading usability studies, and designing databases. The ability to then decide which tools to use to build large, complicated systems based on some shared benchmark reduces the burden of trying to iteratively test through many possible design and methods decisions. When social scientists use NLP, it is much more on the applied side, often adapting code, tools, and algorithms to analyze a particular piece of (textual) data for a topic or question. Applying NLP for social science is not trivial, but there are different expectations and distributions of research labor than with computer science. The strong consensus built in the computer science literature could inadvertently signal to social scientists that these methodological decisions are settled knowledge.

Social scientists are not immune to patterns of methodological monotony, as Abbott observed decades ago in his critique of regression-based quantitative research \cite{abbott1988transcending}. But while sociology's preferences are shaped by intellectual tradition as well as transparency and interpretability, the strong methodological consensus in computer science and NLP can create the illusion of settled knowledge. For outsiders seeking to adopt these tools for their own analyses, determining methodological popularity and convenience from tested justification can be difficult to discern. Within computer science, one of the key reasons why there can be methodological homogenization is because NLP tools are typically developed with an entire computational system in mind that would extract data, transform it, analyze or process it, and then feed it into some other overarching system (e.g., a website). Narrowing the search down to popular models optimized for shared benchmarks can drastically simplify entire sections of a system. Reducing the burden of methodological decision making helps in many different aspects of computer science research, but it can also reinforce convergence around whatever is labeled as the state of the art rather than deeply and time consumingly considering the particulates or their problem space and how less popular models might actually perform better. 

By contrast, social scientists tend to take a more modular approach by adapting methods, code, and techniques to particular datasets to answer specific questions relevant to a discipline. While this does not mean social scientists are completely uninvolved in developing  and adapting NLP tools \cite[see for example][]{jensen2022language,davidson2017automated}, there are stark differences in the expectations about the purpose of a given NLP tool or model between computer and social scientists. Given the tendency in social science to find tools and apply them, a process which would inevitably lead to some kind of online search through general search engines, academic databases, and computer science conference programs, all of which are extremely susceptible to trends reflecting hype and popularity, all of which could create artificial signals about which approach should be used. This also has the double effect of pointing social scientists away from older methods and techniques which could work better with their particular dataset and are more likely to be more transparent and interpretable. Selecting NLP methods is never a purely technical nor even methodological choice, but rather highlights epistemic cultural differences between two broad fields. For these reasons, selecting methods could have much more substantive consequences in the way it could shape all phases of analysis and influence the key takeaways from a given study \cite{stuhler_et_al_chapter}. The consequences of these dynamics can be undervalued as well, and as Figure 1 shows the tradeoffs in method selection between interpretability (favoring simpler, sometimes older methods) and flexibility (favoring more complex, usually state-of-the-art methods) and points to how they could influence social scientific research.

\begin{figure}
    \centering
    \includegraphics[width=\linewidth]{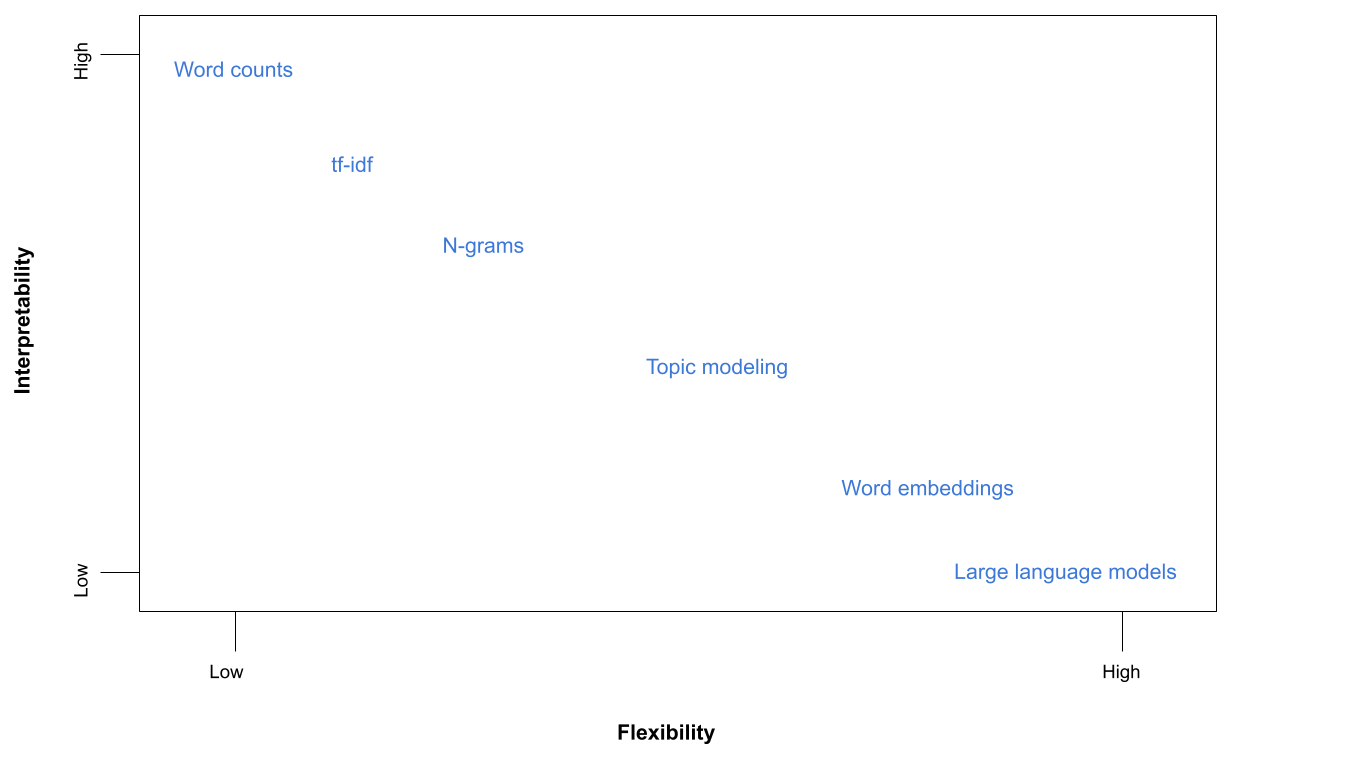}
    \caption{Conceptual diagram of the tradeoffs between interpretability and flexibility in NLP methods used in computational social science. Adapted from \cite{james2013introduction}.}
    \label{fig:enter-label}
\end{figure}

NLP scholars have noted that the state-of-the-art does not always surpass the performance of prior models. As one example, a 2024 paper showed that embedding models (specifically BERT) can match or even exceed the generative capacities of generative models (e.g., GPT), showing how even in settings where newer models build their reputation (like generation) they can still be outpaced by older methods \cite{samuel2024berts}. Another paper found that document topics labeled by generative AI tend to be generic and perform poorly in domain specific corpora \cite{li2025large}. The same has been documented for social science, such as in the case of the 3rd annual Society for Industrial and Organizational Psychology Machine Learning Competition where a ``simple rules based model'' outperformed nearly every machine learning model (51/60), meaning the simpler approach yielded not only high quality outputs but also transparency \cite{harman2023simple}. Sometimes reported findings veer toward borderline absurdity, such as a 2023 paper demonstrating that gzip compression plus k-nearest neighbor classification outperformed BERT on five out-of-distribution datasets \cite{jiang-etal-2023-low}. Simpler and/or older methods therefore do not just offer heightened transparency but also, at times, can perform as well or better than more advanced techniques. The difference though is that social scientists might not have the training or background knowledge to test multiple methods.

A key element in NLP and even ML development in general is the use of massive hidden workforces, referred to as ``ghost workers'' by \cite{gray2019ghost}, who annotate and curate data to be used in model training. Large tech companies can afford to pay thousands of workers to perform these tasks with the goal that these same models will be incorporated into new and existing technology to generate more value and profits. This effects social scientists in two key ways. First, social scientists might not fully realize that tech companies are not going through this extensive labor process for the sake of science but rather because they are creating business tools, not too unlike office equipment like printers and monitors. By consulting the computer science literature or taking courses in computer science departments among graduate students, the fact that social scientists are in no way the intended audience of these tools might lead to misunderstandings about why exactly they are built. They might also not realize that these models, through the extensive use of ghost workers, are the product of extensive top-down mandates where the correct outcome is aligned with financial incentives for companies; this is especially true for generative models. Common refrains explaining limitations in the models, such as bias in the data leading to bias in the outcome, often overlook this fact. 

The second issue is related to ethics and metascience. Ghost work is not easy, especially for those doing things like content moderation. They face a nonstop deluge of hate filled language and imagery, and their job is to meticulously sort through it to determine what should be subject to moderation. But even on the methodological side of research, these tools can present themselves as fully formed, unassailable representations of language and the world. When we use these tools to advance some claim about the world as represented in the model, we might lose sight of how language is constantly changing and full of the kinds of inequality and stratification sociologists tend to examine in their own research. For example, studies have found that models tend to capture features of language usage that is most aligned with dominant dialects and varieties while poorly capturing others \cite{zhang2021sociolectal}. There have been somewhat extreme examples of this tendency, such as models injecting English language syntax into non-English text as well reproducing state propaganda when prompted in Chinese rather than English \cite{waight2024propaganda}. Fully considering the theoretical considerations of bias, ghost work, and epistemic context are beyond the scope of this paper, but it is nevertheless important to note how selecting NLP models (especially the more sophisticated models) can entangle a study in many different ways.

\section*{The State-of-the-Art Fallacy}
The tradeoffs and tensions between the many modeling decisions that go into a given computational research project can be difficult to navigate. The benchmark system developed by computer scientists further complicates this process due to its tendency to reasonably simplify consequential modeling decisions down to selecting the tool, method, or approach which performed the best on a shared benchmark. Research about the design and construction of new models is also simplified based on considerations of shared benchmarks, including whatever manipulations and abstractions from data that achieve the highest accuracy in predictive tasks. Additional outside forces, such as pressures to demonstrate competency in state-of-the-art methodology could also play a key role not just in influencing modeling decisions but even success on the academic labor market. Heiberger et al. point out that targeted specialization and novel combinations will provide career advantages; in the case of sociology, modeling selections, such as quantitative, positively influence their success in sociology \cite{heiberger2021facets}. Social scientists therefore have incentives to use the state-of-the-art \textit{du jour} rather than make methodological decisions best suited to their data and research questions.

As noted, the structure of computer science research, discourse, hype, and publication also facilitate this kind adoption. The term ``state of the art'' serves as a discourse marker to unify computer scientists around a particular approach or method, but in doing so also bring attention from disciplinary outsiders like social scientists, the media, and the general public. In this way, it is a highly effective discourse marker because it originates in a high status, high credibility space imbued with the evaluations of computer scientists (or at least assumptions thereof) based on their own internal competitions \cite{podolny1993status}. The benchmarking system, while not without its issues, also means that these claims are backed by some degree of standardization. To borrow a term from historian Theodore Porter, describing something as ``state-of-the-art'' implies that a given method or model is a ``technology of distance'' that reflects some numerically based achievement rather than individual judgement and evaluation \cite{porter1996trust}. There are many micro-interactions that accumulate to facilitate this process, and future studies might consider adopting theories from Goffman, Heritage, Schlegoff, and others to examine them, but here we take a macro approach and consider Bourdieu's writing about language and linguistic capital \cite{bourdieu1991language}. The interconnections between capital, language technology, and power, as well as how they interact to shape the adoption of NLP tools for social science, make Bourdieu's perspectives apt in our setting.

Bourdieu’s framework helps illuminate how this linguistic capital can traverse disciplinary borders and shape the adoption of methods. Bourdieu argues (like many linguists) that language is not inherently neutral or static but something that is dynamic and possesses capital. That capital can signal value and authority through means like named language (e.g., English), linguistic styles, and in this case, specific terminology (``state of the art''). Once this particular form of linguistic capital is attached to the term ``state of the art'' and its associations with computer science's competitive benchmarks and high status expertise, it enters into sociological domains not as purely technical means but one loaded with additional layers of meaning. Using state-of-the-art tools and describing them as such signals credibility and success validated by a dominant field that increases perceptions of validity and accuracy. 

Even calling some methods state of the art over others belies this reality by suggesting that some methods are inherently more advanced than others rather than more appropriate. We call this the ``state-of-the-art fallacy'' with the goal of encouraging more thoughtful adoption of these tools that become anointed in computational research but might not be the best method to apply in any given social science research question. Simply, the fallacy is the belief that the state of the art for computer science will also be the best tool for any/every sociological question. Without having extensive knowledge of the ways that computer science operates a field, adopting the fallacy is actually a reasonable outcome. Particular methods are marked as ``state of the art'' through the epistemic culture of computer science, and then through regular academic discourse (along with discourse by the media and even just everyday conversation) these methods become more widely known and adopted. But should social scientists therefore assume that the state-of-the-art will work best for their specific research question? On principle we argue that this should not be the case, in our empirical examples we demonstrate why this is the case. 

In no way are we discouraging sociologists from incorporating the full scope of computational methods into their methodological repetoires, nor do we have any papers, theories, findings, or claims in mind as exemplar violations of the fallacy. In fact, there is a reasonable argument to be made that computer scientists themselves fall victim to this fallacy much more often than sociologists\footnote{For example, in a May 2023 virtual mentorship session for the Association of Computational Linguistics (ACL), students and panelists discussed these and similar types of trends, including blunt questions about the value of studying NLP in light of LLMs (\href{https://www.youtube.com/watch?v=g59P3YiqRxU}{Link to video session}).}. Rather, our aim is to provide methodological rationale for why sociologists should not use the current state-of-the-art from computer science and NLP dictate the methods we use in our own research. Doing so would have the inadvertent effect of allowing computer science dictate the types of studies sociologists undertake, a prospect that could limit the sociological imagination of this particular sub-field \cite{stuhler_et_al_chapter,mills1959sociological}. Computational sociology (in the modern digital age) is still nascent, so we hope the way we frame this issue can help individual scientists consider tradeoffs in the methods they select as well as more generally to make a case that we should not just do whatever the computer scientists do. To illustrate this point, we provide an empirical example using computational text analysis and a supervised machine learning (i.e., predictive) framework.

\section*{Empirical Demonstration of the State-of-the-Art Fallacy}
We demonstrate the state-of-the-art fallacy through two sets of analyses on a large dataset of college admissions essays. This type of data is ideal because it exemplifies the kind of important social data that is now more accessible for large-scale analysis than before the advent of NLP tools for social science. Prior analyses also used supervised machine learning approaches and NLP tools (e.g., predict SAT scores, household income, and other applicant characteristics using essay features as predictors), enabling us in this paper to directly compare methods and results. Although most social science does not rely on predictive results, there are many reasons why an accurate model would be useful. For example, new work on methods for causal inference have been adapting machine learning tools due to their ability to make flexible predictions \cite{koch2024primer,brand2021uncovering}, and being able to use text to increase performance on things like matching would be a boon. The abundance of new forms of data has also led to suggestions that social scientists more formally incorporate prediction for hypothesis testing \cite{van2025metatheoretical}. The goal of our examples is to show that if an analyst were to attempt to implement what van Loon calls predictability hypothesizing (where differences in prediction outcomes is the primary metric), there are serious consequences and tradeoffs that must be considered. Importantly, when classification algorithms and NLP techniques are labeled as the state of the art, they are based on prediction tasks, meaning that naively we would assume that they should perform better.  

The first set of analyses compare different NLP methods used to generate features from the essay text and different prediction algorithms. The original work used machine learning but relatively older, simpler methods compared to what was considered to be the state-of-the-art at the time of writing. The goal of this analysis is to imagine the kinds of outcomes that could emerge from assuming that more sophisticated models were going to work better than older models. The second set of analyses consider more recent developments with generative AI that is able to simulate human behaviors (including writing). Generative AI is the current state of the art, and here we consider the dynamics between social identity and synthetic text. Next, we discuss the original corpus of data and each of the studies in more detail.

\subsection*{Data}
We use the same raw essays analyzed by (Author) as well as the computed outputs that are publicly available on \textit{Redacted}. The essays were submitted to the University of California (UC) during the 2015-2016 and 2016-2017 academic years\footnote{The analyses in study one were done in 2020 before the data-use agreement expired. The original essays have since been destroyed.}. UC applicants fill out one application and then apply to whichever campuses they choose. Practically, this means applicants are unable to tailor their essays to particular campuses (e.g., Santa Cruz or Berkeley), making it reasonable to compare essays and applicants. Each applicant had to write four essays (each with a maximum length of 250 words); the original analyses combined them into one ``merged'' document for modeling. The data were used in multiple papers which highlighted the ways that admissions essays had similar issues of bias as other elements of the college application, such as SAT scores and income. For our second study, we use popular LLMs to generate text based on the same essay prompts (in the same proportions selected by applicants). Similar to the ``merged'' documents, we also merge the AI generated essays for analysis. The synthetic text came from GPT-3.5, GPT-4o, Mistral Small (v3), and Claude 3 Haiku. These were selected due to their popularity (especially the GPT models) and ease of use due to their similar chat interfaces. See Table \ref{tab:desc_stats} for more details. 



\begin{table}[h!]
    \centering
    \begin{tabular}{l|c}
       Variable  & Value \\
       \bottomrule
        Essays & 238,892 \\
        LLM generated essays & 101,076 \\ 
        Applicants & 59,723 \\
        Topics & 70 \\
        LIWC features & 89 \\
        Mean and median SAT & 1209; 1210 \\
        Proportion first-gen & 30,210 (not first-gen); 29,513 (first-gen) \\
        Mean and Median GPA & 3.75; 3.81 \\
        School type & 51,232 \% public; 8,488 \% private \\
    \end{tabular}
    \caption{Descriptive statistics of essays and applicant characteristics.}
    \label{tab:desc_stats}
\end{table}

In the original analysis from (Author), they describe using correlated topic modeling \cite{blei2006correlated} and the 2015 edition of the Linguistic Inquiry and Word Count package (LIWC; \cite{pennebaker2015development}). Correlated topic modeling is a 2006 extension of the original 2003 topic modeling paper describing latent Dirichlet allocation \cite{blei2003latent}. As the name suggests, correlated topic modeling differs from traditional topic modeling because it assumes that some topics will be correlated with each other, an assumption not held with latent Dirichlet allocation. This made it ideal because in documents like personal statements, it is likely that different topics will be correlated rather than none at all. The original paper used the ldatuning package in R to determine a suitable number of topics. The package works by comparing models containing different numbers of topics and then suggesting an optimal number of topics based on the average of the results from four different algorithms. \cite{nikita2016select,cao2009density,griffiths2004finding,arun2010finding,deveaud2014accurate}. In this case, they used 70 topics; because topics sum to unity, they removed the topic that was least correlated with each dependent variable (SAT score; household income) to prevent perfect multicollinearity \cite{pescosolido1986social}. Beyond the goal of generating predictors, this approach generated insights into the content of the essays and how it varied along different characteristics of the applicants. 

LIWC is a popular package for computational text analysis that was used in another large scale analysis of college admissions essays \cite{pennebaker2014small}. In that study, the authors found that LIWC features of admissions essays were strong predictors of the final graduating GPA among applicants who eventually enrolled. LIWC is a dictionary method that works by taking counts of particular words and types of words (e.g., articles and pronouns) and cross-referencing them with an external dictionary validated through psycholinguistic research. The package then generates numerical values for each of the features. The original study used 89 LIWC features (all of them excluding the ``Dash'' feature, which was removed for formatting reasons) in their model. The full model in the paper used all of the topics and all of the LIWC features as predictors in a linear regression model that incorporated 10-fold cross validation as an out-of-sample strategy. The final values they reported were the adjusted $R^2$ : \[
    R_{\text{adj}}^2 = 1 - \frac{\smash{\raisebox{0.5ex}{\(\text{SS}_{\text{res}} / \text{df}_{\text{res}}\)}}}{\smash{\raisebox{-0.5ex}{\(\text{SS}_{\text{tot}} / \text{df}_{\text{tot}}\)}}} \]

and the root mean squared error (RMSE):

\[ \text{RMSE} = \sqrt{\frac{1}{n} \sum_{i=1}^n \left( y_i - \hat{y}_i \right)^2} \]

The adjusted $R^2$ is a statistic describing how much of the variance in the dependent variable can be explained by the mode. The ``adjustment'' here (adding degrees of freedom to the usual $R^2$ formula) addresses the limitation that adding more predictors will always increase the $R^2$. The RMSE is a popular metric used in machine learning to describe how well a model can predict out-of-sample outcomes. In this case, the original paper reports the RMSE for models predicting an applicant's SAT score given their essay features. The final adjusted $R^2$ originally reported for explaining variation in SAT scores using CTM was 0.486, and the final RMSE was 124.87 points (less than one standard deviation for the entire test). Combining CTM and LIWC improved the $R^2$ to 0.526. These are the benchmarks we will use to compare state-of-the-art methods from the time that the original paper was written.

Study two uses other information about the applicants and essays generated from AI using the same essay prompts the applicants used. Specifically, we use binary information about the applicants: parental education level (which we label as being first-gen or not depending on whether or not one of their parents completed a college degree); whether the applicant was above or below the median GPA for the applicant pool; and the type of school the applicant attended (binarized to public or private). We also use essays generated from popular LLM models and generate their LIWC features. 

\section*{Study 1: Comparing Other NLP Tools}
In study one, we compare the results from the original study with analyses that use more sophisticated NLP and prediction techniques that have become popular in computer science and sociology (even in the advent of generative AI). Previous work with this dataset focused on predicting SAT score from essay content but assumed a linear relationship between topic prevalence, LIWC features, and SAT score. Additionally, this work did not leverage state-of-the-art machine learning and NLP practices, including flexible models, word embeddings, contextual embeddings, and neural architectures for sequence data. As such, the first study of the current paper focuses on replicating or surpassing previous results, while using more flexible models that allow for relaxing the linearity assumption and changing some modeling decisions present in previous work. Specifically we address three separate, plausible critiques of the original paper: (1) The failure to leverage word embeddings, (2) The approach of correlated topic modeling, and (3) the appropriateness of the linear model.



\subsubsection*{Failure to Leverage Word Embeddings} 
Word embeddings (also commonly referred to as ``word vectors'') have seen increased use in applications involving text because of their ability to encode semantic information into high dimensional spaces and their good performance on common NLP tasks, such as analogy completion. Our analysis makes use of three different types of word embeddings: GloVe \cite{pennington2014glove}, Word2Vec \cite{mikolov2013distributed}, and BERT \cite{devlin2019bert}. All three are pre-trained on large corpora. Importantly, GloVe and Word2Vec use a single vector per word to encode semantic relationships in the embedding space, while BERT allows for variation in the embedding vector depending on the context in which the word is used. We examine the predictive ability of word vector methods and contrast them with topic model based methods. \\

\subsubsection*{Appropriateness of Correlated Topic Modeling}
Because each applicant selects the subset of essays they respond to, a key study design feature is the simplifying choice to concatenate all essays together and fit a single correlated topic model. This may be an insufficiently nuanced approach, as topics may be contextualized by prompt selection (i.e., writing about academic struggle may be weighted differently depending on the prompt). Additionally, the choice of a correlated topic model itself may be fraught. As such, we take two approaches. First, we fit new topic models within each prompt, allowing for information about prompt context and prompt selection decisions to enter our predictive models. 

Second, we implement a neural topic modeling architecture proposed by Grootendorst \cite{grootendorst2022bertopic} based on contextualized embeddings called BERTopic. BERTopic works in three stages: first, it uses the average of BERT-based sentence embeddings to summarize each essay. Second, it clusters essays using these mean sentence embeddings. Finally, it uses within-cluster term frequency, inverse document frequency (TF-IDF) weighting to label each cluster. Each cluster is then assumed to represent a topic and a probability of membership is computed over topics for each essay. These probabilities are analogous to topic loadings in a traditional topic model and used as predictors in subsequent models. 

\subsubsection*{Appropriateness of the Linear Model} 
Linear models may fail to capture appropriate nuance when large numbers of predictors are used, especially in the presence of interactions. For example, it may be the case that writing about overcoming adversity and international travel are each individually associated with higher SAT scores, but writing about them together is associated with lower scores. Our primary tool for addressing this critique is extreme gradient boosting (XGBoost), a tree-based ensemble machine learning algorithm that works well with tabular data, fits quickly, can learn interaction structures, and is resistant to overfitting \cite{chen2016xgboost}. We implement XGBoost liberally across transformations of the essays and compare its performance to linear models in multiple cases.

As we have access to full text of essays, another concern is that linear models of topics loadings or word vectors may be inappropriately sensitive to the way essays are constructed. Because of this, we implement two separate neural approaches to predicting SAT score from raw text. The first is a bidirectional long short-term memory (LSTM) model that treats essay text as a sequence of words. The second is a one dimensional convolutional neural network (CNN) that looks for patterns among the relative placement of words in text. Similarly, Chen and Guestrin demonstrated that XGBoost, despite being relatively dated, remains effective in handling large textual datasets efficiently. These examples underscore the importance of not overlooking tried-and-tested methods in the pursuit of innovation. 

\subsection*{Study 1: Methods}


\subsection*{Tree-based approach}

For each student, we generated GloVe-based embeddings using the average embedded response for each prompt. Using all 300 embedded dimensions, we trained an xgboost model while varying the learning rate[-5,. 01], tree depth [5,15], and the number of iterations. The model was trained using 50 trees using a grid-based sampling technique, and we tuned model performance on a validation sample using early stopping .

\subsection*{Other Topic Models}
We also experimented with separate evaluations where we trained topics models independently by essay prompt or jointly allowing topics to vary by prompt. We also tested ensemble approaches that took the average of each student's prediction based on each individual essay.

\subsubsection*{1D-CNNs for Textual Data}

We follow the approach of \cite{kim2014convolutional}, but instead apply the method to prediction of continuous SAT scores instead of categorical classifications. Individual words in concatenated essays are first projected into $\mathbb{R}^{300}$ using pre-trained Word2Vec embeddings \cite[]{mikolov2013distributed}. A 1D-CNN used for text classification will apply \textit{filters} to sliding windows of $k$ words at a time and produce weighted sums of the individual components of the word vectors, learning optimal weights to minimize the mean squared error in the prediction of SAT scores. Each individual filter contains $300k$ learned parameters. Because the CNN treats the words in an essay as a sequence, it respects ordering, distance, and multi-word groupings in ways that bag-of-words and embedding averaging methods do not. To combat potential overfitting, we limited the number of individual filters and employed dropout at the whole filter level.



\subsection*{Study 1: Results}

\subsubsection*{Model Selection} 


For the 1D-CNN, models were compared over number of training epochs, learning rates, dropout probabilities, and number and width of convolutional filters. The best performing model we observed was comprised of $24$ total filters, three each of $k \in \{3, 5, 10, 20, 50, 100, 150, 200\}$ (resulting in $484,200$ learned parameters) with a learning rate of $\lambda = 0.05$, a dropout probability of $0.2$. Performance plateaued after 100 epochs of training, producing an RMSE of about $143$ points in held out testing data.


\subsubsection*{Comparison of Results}

\begin{figure}[h!]
    \centering
    \includegraphics[width=\linewidth]{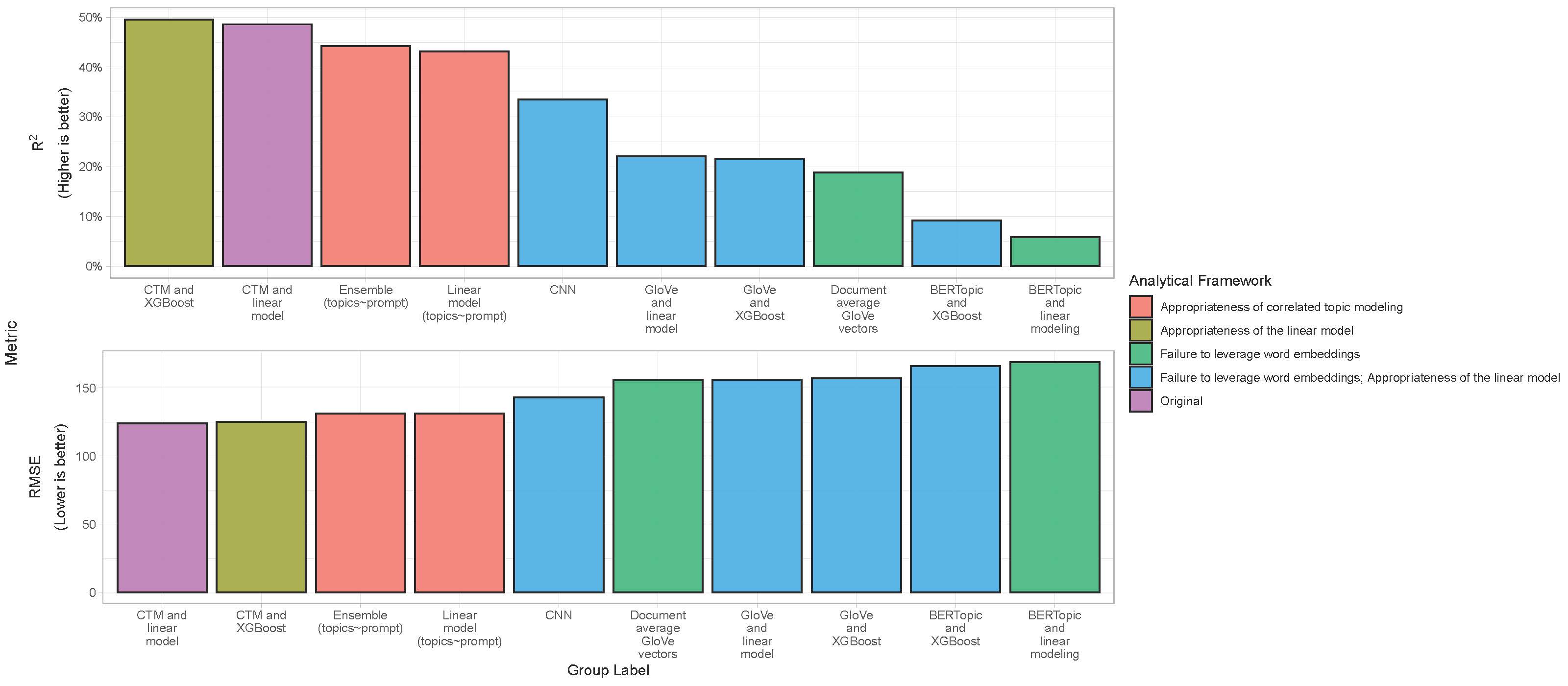}
    \caption{CTM was as good or better than other methods tested. CTM and XGBoost had a higher $R^2$ but lower RMSE than CTM and linear modeling.}
    \label{fig:study_one_results}
\end{figure}

Figure \ref{fig:study_one_results} presents the primary findings for study one, namely that CTM features were stronger predictors. It was also the case that XGBoost and CTM generated a higher $R^2$ but lower RMSE (0.495 vs. 0.486 and 125 vs 124 respectively), meaning it had slightly higher average error than out-of-sample linear modeling though explained approximately 1\% more of the variance. Among the other methods tested, the $R^2$ ranged from 0.44 (Ensemble topics per prompt) to 0.05 (BERTopic and linear modeling). The weaker performance of the embedding approaches was somewhat somewhat surprising given that each of these methods relied on more predictors (which always increase $R^2$) but still failed to surpass the original approach. The RMSE is also subject to this particular limitation (though the RMSE always goes down instead), and the basic results were still the same. These results suggest that the increase of complexity and the use of popular, state-of-the-art models (at the time of writing and analysis) do not always improve the predictive performance, even though theoretically they could by default given their high dimensionality. It is possible that different random seeds for the XGBoost model could have tilted the final outcome in another direction as well \cite{picard2021torch}. XGBoost is a relatively interpretable machine learning method compared to deep learning and other black box tools, but it is far less so than traditional linear modeling. Computer scientists might see the 1\% increase in performance on $R^2$ and declare XGBoost to be the new state of the art, and in benchmark settings they would not be wrong. But social scientists should be wary of adopting similar mentalities given the relatively small increase in performance in exchange for much less relative interpretability.

It would be easy to imagine someone testing the more complex methods, seeing the relatively poor performance, and halting further inquiry (especially with the popular BERTopic approach). The meta-point of study is that methodological plurality can help us identify stronger methods, even if they are not the current state of the art. Unfortunately, for computational social scientists, these decisions will likely require more time testing different models with different strengths and weaknesses, especially compared to the typically more focused, regression based approaches favored by much of traditional social science. This could also mean going against computer science orthodoxy in ways that could invite critique and scrutiny, even if the strength of the results from using older methods is notable (like they are here).

\section*{Study 2: Silicon Sampling and Authorship Biases}

The second study shifts focus to what is, as of 2022, the current state of the art in AI and NLP: generative AI (specifically LLMs). LLMs are not well suited for the kind of direct comparisons described in the first study, but the transformative potential of generative AI for social science has been noted elsewhere \cite{bail2024can}, therefore we do not discuss it at length here. Methodologically, much of the sociology which incorporates LLMs has focused on the potential for AI models to generate realistic, or at least plausible, data which resemble human subjects data, an approach called ``silicon sampling'' or \textit{in silico} social science \cite{kozlowski_evans_2025,argyle2023out}. LLMs could also be used for information extraction and summarization from text \cite{stuhler2025codebooks}, but to keep Study 1 and 2 at least somewhat comparable we do not spend time on such analyses. With silicon sampling, there are enormous practical and intellectual implications to the way sociologists do research if their state-of-the-art label is backed by strong performance: reducing the costs of running surveys, faster piloting and prototyping, and potentially even opening up new lines of inquiry. Most dramatically, scholars working in this space have floated the idea that, eventually, LLMs could eliminate the need for human subjects in research.


The early results in silicon sampling have been promising in a few key settings but less rosy in others. Studies have shown that AI models can faithfully reproduce results to surveys through their responses, predict future responses (LLMs are temporally bounded by the data used to train them), and map closely to responses broken out by political affiliations (when the models are given political cues in the prompts) \cite{argyle2023out,kozlowski2024silico,kim2023ai}. There are less splashy uses of LLMs in social science, such as high quality data annotation, but these and similar results from the silicon sampling literature are likely to grow in prominence. Other studies have posed various, though not always direct, challenges to silicon sampling and these results. For example, issues related to machine bias and the tendency to mimic Western, Educated, Industrialized, Rich and Democratic (WEIRD) over other people and communities are an issue for LLMs \cite{santurkar2023whose,tao2024cultural}. In our second study, we consider the potential for silicon sampling to produce college admissions essays and whether or not the text they generate shows similar issues of bias or not. We first hypothesize a few reasons why scholars would want to take this approach.



\subsubsection*{Expanding analytical breadth and depth}
There are many kinds of data, including college application materials, that are hard to obtain for social scientific inquiry despite their importance. In our context, admissions essays are the third most important element of the application (after GPA and test scores), but their prominence has risen since the COVID-19 pandemic. In response to social distancing mandates, universities instituted test-optional and test-blind policies, meaning that the relative of weight of the essay might have changed. Then, in 2023 the Supreme Court of the United States banned race-based affirmative action in admissions, though they were careful to note that essays which included descriptions of racial hardship were exempt. These events can and are being studied without examining essays, but setting up a study which includes them would expand its analytical breadth in obvious ways. Without having actual essays to test, using an LLM to generate responses to the same essay prompts as the actual humans could be a convenient workaround.

Advances in statistical methods for conjoint experiments, causal inference with text, and simulations also make the use of text in a given study more analytically tractable \cite{egami2022make,thompson2023they}. For example, imagine an experiment where participants are asked to evaluate hypothetical applicants where the treatment arms are different reflections of real world events (e.g., one condition might not have an essay at all). Soliciting text from many participants can become expensive; even if not prohibitively expensive, it could require time spent securing some source of funding at the expense of timeliness.

\subsubsection*{Each LLM is the state of the art in something}
The benchmark system used in machine learning has been the subject of critique, though not necessarily in a way that would deter sociologists from using them to generate text. In the corporate world of LLM production, the benchmark system has become nearly indistinguishable from the kinds of marketing and advertising claims made in other industries. Each new model is released in a press conference where its strengths are touted, including many claims about test performance and the ways they can claim their model is the state of the art. For social scientists, dubious claims made by corporate entities matter less than the subtext that all of these models work pretty well at a variety of tasks. OpenAI's models, especially in the ChatGPT online interface, are the most popular \cite{zhang_xu_alvero_2024} but their performance relative to others (e.g., Anthropic's Claude models) are not drastically different. ChatGPT set a strong precedent about the format and tendencies that the public would adopt, and in classic organizational fashion other companies followed suit.


\subsubsection*{Ease of use with prompting}
A major reason why LLMs have become so popular is the ease of their use, specifically with the ability to have users write a prompt, a short message containing detailed instructions about what the user wants the model to generate. To run an experiment on the effects of changes in application protocols with respect to the essay, an analyst could simply prompt an AI to generate essays using the same essay prompts as human applicants. Further, to induce the models to write particular types of essays (e.g., essays about racial hardship to connect to the Supreme Court decision on race-based affirmative action), simply editing the prompt should in theory get the model to respond in kind. This could mean prompting the model to specifically write an essay which includes an example of a racial hardship or prompting it to write from the perspective of a racial minority; the latter seems to be effective in survey question responses \cite{argyle2023out}. Sociologists might be susceptible to vibe theorizing and marketing hype in other ways too. Generally, many sociologists do not explicitly consider language in their work, even among those who do computational text analysis (though there are notable exceptions, e.g., \cite{jones2020stereotypical} \cite{jones2020stereotypical}). Some in the AI space have argued that the outputs from LLMs represent the aggregate of human knowledge, text, language, and culture and therefore are the closest reflections to the ``true distribution of language'' we currently have available to us \cite{wang2023large} and can be considered linguistically ``neutral''.  Given that the study of language is not common in sociology (aside from the ethnomethodologists carrying on the traditions of Harold Garfinkel and others) or often even considered, countering perspectives might not be readily available or even known.

\subsection*{Study 2: Methods}
Study 2 considers the scenario where an analyst takes in these and other rationale for using an LLM to generate admissions essays. However, we depart from Study 1 by determining the suitability for LLMs to simply generate sample essays to use for analysis in the kinds of experiments we described above. There are many ways to answer this question, but to maintain some coherence across the studies we take a similar sociodemographic approach but instead focus on other dimensions of the applicants, specifically their GPA, parental education level, and school type. Our goal with Study 2 was to see if using LLM generated essays out-of-the-box could introduce unforeseen confounders that might not be on the top of mind for sociologists. We do this by testing the demographic alignment of AI generated text with respect to important sociodemographic categories. We use the same essay prompts used by the human applicants as prompts for a variety of LLMs to see how well they can be used ``out-of-the-box'' to faithfully simulate patterns in human writing. We test four different models: GPT-3.5, GPT-4o, Mistral Small (v3), and Claude 3 Haiku. The GPT essays were generated for another study (Author), and the Claude and Mistral essays were generated for this study. These are popular, smaller (read cheaper) models that would be accessible to sociologists, allowing us to also interpret the results from the perspective of what might such an analysis look like before ending up in the file drawer. 

Rather than CTM, here we use LIWC features as predictors. The LIWC features are strongly correlated with many different sociodemographic factors, allowing us to extrapolate our findings towards more general questions of AI vs. human writing styles in the contexts of college admissions. Practically, this allowed us to directly compare the human and AI generated essays since LIWC works by taking word counts in each document. We excluded the feature that literally counted the number of words in the essays due to some minor variation in the AI generated essays. After generating the LIWC features for the AI essays, we followed (Author) and trained a machine learning model to predict the sociodemographic characteristics of the authors given their essays. Table \ref{tab:model_accuracy} presents the results. Though these models were not extremely accurate, especially not compared to the numbers typically reported in the machine learning literature, the goal was to test for alignment, and absent extremely \textit{in}accurate models we moved forward with our analysis. Specifically, these models were trained to predict whether or not an essay was written by an applicant with above/below median GPA; parental education level (split by college educated or not); and the type of school they attended (split between public and private). The trained models were then given the AI generated essays to label, providing insight into who the LLMs write like. 

As a robustness check, we also test different splits of the training and testing data for each of the machine learning models. Our first set of models used a more conservative 50\% training, 50\% test split of the human essays to follow recommended procedures laid out by social scientists working with text \cite{egami2022make}. We also tested more skewed splits favored by computer scientists, specifically an 80/20 and a 90/10 train-test split. Importantly, there were only minor differences between the results, and for simplicity we just focus on the results for the 50/50 split. See Tables \ref{tab:study_two_results_1} and \ref{tab:study_two_results_2} for more the other results.

\subsection*{Study 2: Results}
Table \ref{tab:study_two_results} presents the primary results of study two. Across all of the models tested, we corroborate other examinations of generative AI and college admissions essays and find significant sociodemographic alignment out-of-the-box \cite{alvero2024large,lee2025pooralignmentsteerabilitylarge}. For example, a model trained to predict whether or not an applicant's parents attended college predicted that essays written by LLMs were much more likely to also come from those with college educated parents (ranging from 74\% to 90\%). This is despite the distribution of the actual data being roughly split in half. For GPA, another feature of the applicants split down the middle but also salient in essay content and style, the models were mixed. GPT-3.5, Claude 3 Haiku, and Mistral Small wrote much more like students with below median GPA (85\%, 93\&, and 60\% respectively). GPT-4o on the other hand was nearly split perfectly in half. This tendency was somewhat unexpected given the patterns in parental education, but it does show how model selectiong might be especially fateful with LLMs, especially considering that these models could be withdrawn from public use at any time. 

Finally, models the most imbalanced feature of the applicants we considered, the type of school they attended (85\% attended a public high school) predicted that most of the AI generated essays were from private school applicants (ranging from 95\% to 60\%). LLMs in general wrote more like applicants who attended private schools and had college educated parent. However, they also more often wrote like applicants with below median GPA than not (aside from the exception of GPT-4o). Another paper described taking a different approach by also including demographic information (e.g., parental education level, ethnoracial identity, gender) but found no change in their similar results \cite{lee2025pooralignmentsteerabilitylarge}. Using LLMs for this kind of more complex work (compared to survey response simulations) would introduce biases that would be difficult to disentangle.

\begin{table}[h!]
  \centering
  \caption{Descriptive Statistics by Data Source and Model}
  \label{tab:study_two_results}
  \begin{tabular}{lccccc}
    \toprule
    \textbf{Category} & \textbf{Human} & \textbf{GPT-3.5} & \textbf{GPT-4o} & \textbf{Claude} & \textbf{Mistral} \\
    \midrule
    
    
    \textbf{GPA Median} \\
    \quad Low & 29,604\textsuperscript{a}  & 11,227 & 5,083 & 931 & 609\\
    \quad & (50.3\%) & (85.5\%) & (50.1\%) & (93.1\%) & (60.9\%)\\
    \quad High & 29,272  & 1,899 & 5,060 & 69 & 391 \\
    \quad & (49.7\%) & (14.5\%) & (49.9\%) & (6.9\%) & (39.1\%)\\
    \addlinespace
    
    \textbf{Parents' Education} \\
    \quad First Gen. & 28,616 & 1,266 & 2,362 & 122 & 260 \\
    \quad & (48.6\%) & (9.6\%) & (23.3\%) &  (12.2\%) & (26.0\%) \\
    \quad Continuing Gen. & 30,210 & 11,860 & 7,781 & 878 & 740 \\
    \quad & (51.4\%) & (90.4\%) & (76.7\%) & (87.8\%) & (74.0\%)\\
    \addlinespace
    
    \textbf{School Type} \\
    \quad Public & 51,232 & 660 & 995 & 392 & 161 \\
    \quad & (85.8\%) & (5.0\%) & (9.8\%) & (39.2\%) & (16.1\%) \\
    \quad Private & 8,488 & 12,466 & 9,148 & 608 & 839 \\
    \quad & (14.2\%) & (95.0\%) & (90.2\%) & (60.8\%) & (83.9\%) \\
    \addlinespace
    
     \bottomrule
  \end{tabular}
  
\vspace{0.5em}
{\footnotesize 
\textit{Note:} \textsuperscript{a}637 median as ``Low''}
\end{table}

\section*{Discussion}

This paper introduces the ``state-of-the-art fallacy'', the argument that using the tools deemed to be state of the art by computer science do not always yield the best results; we focus on NLP. While this approach might make sense for computer scientists who are considering NLP tools in the context of larger computational systems, social scientists predominantly use NLP in applied settings, making the process of selecting a method much more consequential to the integrity of a given research question. We provide two empirical demonstrations of how the fallacy might play out through a comparison of NLP methods ranging from more and less state of the art (at the time of writing) and an analysis showing how some approaches might create more covert issues. The first study found that CTM and linear modeling outperformed a variety of both NLP and prediction methods, even those that were considered to be the state-of-the-art at the time the original study was published. In fact, the adjusted $R^2$ for the lowest performing approach, BERTopic with XGBoost, was nearly ten times lower than the original approach from the 2021 paper (0.058 and 0.486 respectively). The same approach also reported an RMSE 1.36 times higher than the original findings. The second study focused on the current state of the art, generative AI, and found problems related to alignment where the text outputs most closely resembled patterns from particular subgroups. For example, after training a model to predict the parental education level of students given their admissions essays, the same model predicted AI generated essays from different models to be much more similar to those written by applicants who had at least one parent graduate from college. The same was true for even highly imbalanced data like school type (public or private).


Variation in the results due to differences in methodology is itself not surprising, but the nature of the variation is important. $R^2$ in a model will increase when you add more independent variables, yet despite that the original analyses using a lower dimensional representation of the essays generated from older methods outperformed even the much higher dimension word embedding approaches. The degree of the discrepancies in findings were so strong that it could discourage researchers to pursue further lines of inquiry using NLP and this particular dataset. Additional risks could arise in scenarios like study two where the state-of-the-art methods seem to be working. Given a prompt provided by a user, generative AI \textit{will} respond and produce something, giving the impression that the models are working as intended. While this can sometimes be the case, as seen in recent studies examining AI generated survey responses \cite{argyle2023out,kim2023ai,kozlowski2024silico}, studies involving text outputs from LLMs (which could then adapt NLP tools) are more complicated. Sociology in particular is at hazard for the kinds of sociolinguistic bias we report in study two, where text outputs are much more aligned with groups from higher social status backgrounds. Decades of sociolinguistic theory and empirical research (not to mention related work from linguistic anthropology and other disciplines studying language and society) has argued that sociolinguistic variation along sociodemographic dimensions is central to essentially all aspects of language, such as dialect, perception, semantics, pragmatics, and vocabulary. Using LLMs to generate linguistic artifacts for sociological inquiry might benefit from drawing upon this literature as a starting point. Study two articulates the trouble with ignoring sociolinguistic patterning in text in favor of the belief that state-of-the-art, pre-trained text models do not reflect these same patterns.



Following trends in computer science determines what we ask and how we answer, sometimes to deleterious results. While our analyses demonstrate a much more focused instance of these dynamics with college admissions essays, others have reached similar conclusions at larger scales by pointing out how this approach to doing social science could limit our sociological imagination \cite{stuhler_et_al_chapter}. The pressure and influence of computational hype, but there are a few ways that sociology could buttress itself when it comes to applying NLP to our research questions, obviously including our argument that we should not fall for the state-of-the-art fallacy. First, if we accept that these tools were not designed with our questions, theories, or perspectives in mind, sociologists could spend more time formalizing the connections between NLP methods and sociology. For example, Stoltz and Taylor have developed thorough, robust methodological frameworks and tools for sociologists to connect cultural and cognitive sociology to word embedding models \cite{taylor2020concept,stoltz2024mapping}. Stewart and his collaborators (many of whom are political scientists) has likewise done significant work connecting causal inference methods used often in social science to NLP \cite{roberts2014structural,rodriguez2023embedding}. Social network analysis emerged as popular methodology in similar ways. 

Another way that sociologists could ensure we do not fall victim to the fallacy is through plurality. Unlike other social sciences where there is a strict focus on identifying and isolating exact social mechanisms and their effects (e.g., economics), sociologists are, perhaps infamously, broad in their approach to question answering. Sociology also places a high value on causal research, but it also places high value on interesting descriptive work, qualitative work, and purely theoretical work. Our empirical analyses, especially study one, took this approach implicitly by testing many different approaches and methods and comparing the outcomes. Leveraging computational processing as the modern innovation (rather than a specific NLP method) and the ability to try many different methods and models has some proponents in sociology. Young took this approach in an article about economic growth and a methods paper where he and a co-author ran ``9 billion regressions'' \cite{young2009model,munoz2018we}. Sotoudeh and DiMaggio make similar arguments about applying grid searches (where different models and specifications are tested and compared to each other), though in their case they were not focusing specifically on NLP \cite{sotoudeh2023coping}. This approach might seem like cherry picking, but with NLP we are almost always going to be doing some level of cherry picking since the strong tendency of this work is to apply an NLP tool rather than build one ourselves. And the critique of cherry picking seems relatively minor considering the alternatives of forcing one method to predominate text analysis \textit{a la} linear regression \cite{abbott1988transcending} or ignoring the increasing saturation of text in our modern social world. In light of these less appealing outcomes, testing and comparing models would be a relatively small price to pay.

A reasonable interpretation of our argument is that as a rule we should avoid using whatever is the state of the art, but that is not our position. In fact, we recognize that there will be instances where the state of the art will turn out to be the best tool for a given study. LLMs and generative AI can also be used for other media that traditional machine learning methods would struggle with, such as audio, video, and image. Although our primary concern is the state of the art fallacy, we must not fall into a state-of-the-art-fallacy fallacy and assume that the latest methods will never work for our own aims.

\subsection*{Limitations}
While our modeling approach is varied, it is by no means exhaustive. It is quite possible that we were just on the cusp of trying the ``right'' model, or a future advance will outperform any model presented here. Additionally, we make no effort to assert that correlated topic models and linear regression are the best way to model similar data. What we do assert, however, is the importance of establishing quality performance baselines and careful thought about the flexibility-interpretability tradeoff. The baseline leads to interpretations that can be easily communicated to a non-techincal audience (students who write about participating in debate tend to get higher SAT scores). While interpretable machine learning is an active area of research, explaining the decisions made by neural architectures is still challenging.

Most importantly, the good performance and simple structure of the baseline model suggest a hypothesis that admissions officers could develop a similar schema when they read essays. More experimental work is needed, but evidence supporting this hypothesis could imply that essay content carries enough information to undermine the goals of test optional and test blind admissions policies.

\textit{Redacted}

\newpage
\bibliographystyle{unsrt}  
\bibliography{references}  

\newpage
\appendix
\section*{Appendix}



**Appendix**

| University | Course offered? | Course(s) | NLP/Text Analysis? |
| --- | --- | --- | --- |
| Harvard | Yes | SOCIOL 2272 | Information not publicly available |
| Princeton | Yes | SOC 592, 596 | Yes |
| Stanford | Yes | SOC 281 | Yes (crosslisted, offered by the Department of Psychology) |
| UC Berkeley | Yes | SOCIOL 273L, 273M | Yes |
| Michigan | No | | |
| UCLA | Yes | SOCIOL 208c | Yes |
| Northwestern | Yes | SOCIOL 476 | Yes |
| Chicago | Yes | SOCI 20602 | Yes |
| UNC Chapel Hill | No | | |
| Wisconsin | No | | |
| Columbia | No | | |
| NYU | Yes* | | |
| Penn | No | | |
| Cornell | Yes | SOC 6610, 6130 | Yes |
| Duke | Yes | SOCIOL 703, 367S* | Yes |
| Indiana | No† | | |
| Texas | No | | |
| Ohio State | No | | |
| Yale | Yes | SOCY 1283a*, SOCY 3203a* | Yes |
| Brown | Yes | SOC 2071 S01 | No |
| UC Irvine | Yes | SOCIOL 212, 229, 281 | No |
| Minnesota | No | | |
| Washington | Yes | SOC 506, 537 | No |
| Johns Hopkins | Yes | AS 230.618 | No |
| Penn State | Yes | SOC 580 | No |
| Maryland | Yes | SOCY618 | No |
| UMass Amherst | No | | |
| CUNY | Yes | Soc. 82301 | No |
| Rice | Yes | SOCI 528 | No |
| Arizona | Yes | SOC 501, 526, 561 | Yes |
| Notre Dame | Yes | SOC 63905 | Yes |

Table A1: Graduate level computational sociology course offerings with a focus on methods training and development at major U.S. sociology departments. Departments represent the top 30 programs (including ties) according to the U.S. News World and Rankings. Many of these courses also state that they have no coding requirements/expectations, e.g., AS 230.618. Most of the "no" departments offer classes which focus on social network analysis. Others focused on programming/coding. CUNY and Rice's offerings focused exclusively on GIS methods. One (at Northwestern) focused exclusively on NLP. Some programs offer undergraduate classes and/or classes studying computation and computers as a topic, though such classes were excluded.
* Unclear how many, but department website indicates courses offered occasionally.
† No public information. A search on Coursicle, a website which hosts course description and syllabus information, did not have computational courses in the department.



\begin{table}[h!]
  \centering
  \caption{Descriptive Statistics by Data Source and Model 80/20}
  \label{tab:study_two_results_1}
  \begin{tabular}{lccccc}
    \toprule
    \textbf{Category} & \textbf{Human} & \textbf{GPT-3.5} & \textbf{GPT-4.0} & \textbf{Claude} & \textbf{Mistral} \\
    \midrule
    
    \textbf{GPA Median} \\
    \quad Low & 29,604\textsuperscript{a} & 10,865 & 4,987 & 907 & 546 \\
    \quad  & (50.3\%) & (82.8\%) & (49.2\%) & (90.7\%) & (54.6\%) \\
    \quad High & 29,272 & 2,261 & 5,156 & 93 & 454 \\
    \quad  & (49.7\%) & (17.2\%) & (50.8\%) & (9.3\%) & (45.4\%) \\
    \addlinespace
    
    \textbf{Parents' Education} \\
    \quad First Gen. & 28,616 & 1,423 & 2,091 & 111 & 255 \\
    \quad  & (48.6\%) & (10.8\%) & (20.6\%) & (11.1\%) & (25.5\%) \\
    \quad Continuing Gen. & 30,210 & 11,703 & 8,052 & 889 & 745 \\
    \quad  & (51.4\%) & (89.2\%) & (79.4\%) & (88.9\%) & (74.5\%) \\
    \addlinespace
    
    \textbf{School Type} \\
    \quad Public & 51,232 & 323 & 737 & 339 & 103 \\
    \quad  & (85.8\%) & (2.5\%) & (7.3\%) & (33.9\%) & (10.3\%) \\
    \quad Private & 8,488 & 12,803 & 9,406 & 661 & 897 \\
    \quad  & (14.2\%) & (97.5\%) & (92.7\%) & (66.1\%) & (89.7\%) \\
    
    \bottomrule
  \end{tabular}
  
\vspace{0.5em}
{\footnotesize 
\textit{Notes:} \textsuperscript{a}637 median as ``Low''}
\end{table}

\begin{table}[h!]
  \centering
  \caption{Descriptive Statistics by Data Source and Model 90/10}
  \label{tab:study_two_results_2}
  \begin{tabular}{lccccc}
    \toprule
    \textbf{Category} & \textbf{Human} & \textbf{GPT-3.5} & \textbf{GPT-4.0} & \textbf{Claude} & \textbf{Mistral} \\
    \midrule
    
    \textbf{GPA Median} \\
    \quad Low & 29,604\textsuperscript{a} & 11,199 & 5,237 & 924 & 591 \\
    \quad  & (50.3\%) & (85.3\%) & (51.6\%) & (92.4\%) & (59.1\%) \\
    \quad High & 29,272 & 1,927 & 4,906 & 76 & 409 \\
    \quad  & (49.7\%) & (14.7\%) & (48.4\%) & (7.6\%) & (40.9\%) \\
    \addlinespace
    
    \textbf{Parents' Education} \\
    \quad First Gen. & 28,616 & 1,165 & 1,914 & 110 & 220 \\
    \quad  & (48.6\%) & (8.9\%) & (18.9\%) & (11.0\%) & (22.0\%) \\
    \quad Continuing Gen. & 30,210 & 11,961 & 8,229 & 890 & 780 \\
    \quad  & (51.4\%) & (91.1\%) & (81.1\%) & (89.0\%) & (78.0\%) \\
    \addlinespace
    
    \textbf{School Type} \\
    \quad Public & 51,232 & 392 & 844 & 358 & 114 \\
    \quad  & (85.8\%) & (3.0\%) & (8.3\%) & (35.8\%) & (11.4\%) \\
    \quad Private & 8,488 & 12,734 & 9,299 & 642 & 886 \\
    \quad  & (14.2\%) & (97.0\%) & (91.7\%) & (64.2\%) & (88.6\%) \\
    
    \bottomrule
  \end{tabular}
  
\vspace{0.5em}
{\footnotesize 
\textit{Notes:} \textsuperscript{a}637 median as ``Low''}
\end{table}

\begin{table}[htbp]
  \centering
  \caption{Prediction accuracies}
  \label{tab:model_accuracy}
  \begin{tabular}{lc}
    \toprule
    \textbf{} & \textbf{Model Accuracy} \\
    \midrule
    GPA Median & 65.11\% \\
    Parents' Edu. Level & 70.50\% \\
    School Type & 63.22\% \\
    \bottomrule
  \end{tabular}
\end{table}

\end{document}